\documentclass{article}

\usepackage{arxiv}

\usepackage[utf8]{inputenc} 
\usepackage[T1]{fontenc}    
\usepackage{hyperref}       
\usepackage{url}            
\usepackage{booktabs}       
\usepackage{nicefrac}       
\usepackage{microtype}      
\usepackage{lipsum}

\usepackage{cite}
\usepackage{amsmath,amssymb,amsfonts,amsbsy}
\usepackage{algorithmic}
\usepackage{graphicx,xcolor}
\usepackage{textcomp}
\usepackage{xcolor}
\usepackage{subfigure} 
\usepackage{adjustbox}
\usepackage{comment}

\newcommand{\hlr}[1]{\textcolor{red}{#1}}

\newcommand{\bs}[1]{\boldsymbol{#1}}

\begin{document}

\title{Mu-suppression detection in motor imagery electroencephalographic signals
using the generalized extreme value distribution} 

\author{
  Antonio Quintero-Rinc\'on, Carlos D'Giano\\
  Centro Integral de Epilepsia y Telemetr\'ia\\
  Fundaci\'on Lucha contra las Enfermedades Neurol\'ogicas Infantiles (FLENI)\\
  Buenos Aires, Argentina \\
  \texttt{tonioquintero@ieee.org, cdigiano@fleni.org.ar } \\
   \And
 Hadj Batatia \\
  IRIT - INPT\\
  University of Toulouse\\
  Toulouse, France \\
  \texttt{hadj.batatia@inp-toulouse.fr } \\
}

\maketitle

\begin{abstract}
This paper deals with the detection of mu-suppression from electroencephalographic (EEG) signals in brain-computer interface (BCI). For this purpose, an efficient algorithm is proposed based on a statistical model and a linear classifier. Precisely, the generalized extreme value distribution (GEV) is proposed to represent the power spectrum density of the EEG signal in the central motor cortex. The associated three parameters are estimated using the maximum likelihood method. Based on these parameters, a simple and efficient linear classifier was designed to classify three types of events: imagery, movement, and resting. Preliminary results show that the proposed statistical model can be used in order to detect precisely the mu-suppression and  distinguish  different EEG events, with very good classification accuracy.
\end{abstract}

\keywords{Motor imagery \and Mu-suppression \and Generalized extreme value \and Electroencephalography \and Brain-computer interface}

\section{Introduction}
\label{sec:intro}
Electroencephalograms (EEG) are a non-invasive longstanding medical modality that measures the brain's activity by recording the electromagnetic field at the scalp. Since its creation, EEG has played a fundamental role in understanding several major neurological disorders, by analyzing their manifestation into brain rhythms. For example, the study of deceases such as depression, age-related cognitive deterioration, epilepsy, anxiety disorders and subnormal brain development in children have benefited from this technology. The typical brain rhythms are distinguished by their different frequency ranges, called delta ($\delta$) within the range 0.5 to 4Hz, theta ($\theta$) within the range 4 to 7.5Hz, alpha ($\alpha$) within the range 8 to 13Hz, beta ($\beta$) within the range 14 to 30Hz,  and gamma ($\gamma$) within the range 30 to 64Hz. In this study, we focus on the brain rhythm called mu ($\mu$) within the range 7.5 to 11.5Hz. 
Mu-waves are considered to emerge naturally  and may convey information about what the functioning of brain hierarchies \cite{Friston2019}.
According to \cite{Niedermeyer2010}, there exist three historical theoretical hypotheses to explaining the mu-brain rhythm: i) the neuronal hyperexcitability related to the rolandic cortex; ii) the superficial cortical inhibition explaining its suppression with motor activity; and iii) the somatosensory cortical idling, related to the afference-dependent phenomenon. This study considers the second hypothesis, as the mu-rhythm relates strongly to the sensorimotor cortex and associated areas, in particular, the changes in the bilateral brain activities subject to physical and imaginary movements \cite{Sanei2013}.
Based on the same consideration, this rhythm has been studied in brain-computer interface (BCI) \cite{Arroyo1993,Pfurtscheller1994}.
The underlying idea of BCI is to supply communication and control  of devices through the monitoring of  brain activity, by using EEG channels.

The generalized extreme value (GEV) distribution is a family that includes continuous probability distributions obtained as the limit of maxima of a sequence of independent and identically distributed random variables \cite{Castillo2004}. This distribution has been used in \cite{Gaspard2014} to detect interictal spikes in epileptic EEG signals, using time-frequency properties. The underlying idea was to identify strong outliers using GEV to model normalized EEG data. In \cite{Nazmia2015}, showed that both EEG and MEG signals can be correctly modeled using GEV distribution. Luca et al. \cite{Luca2014} used an unsupervised method to detect hyper-motor epileptic seizures, where GEV was applied for extracting maxima in EEG signals, using multivariate kernel density.
In \cite{Ueda2009}, GEV was used to assess characteristics of Alzheimer's disease using EEG signals, where the variance of the power of each frequency were used to derive an index of neuronal abnormality. For applications in other biomedical signals see \cite{Roberts2000}.  

The purpose of this paper is to present a novel and rapid algorithm for detecting  mu-suppression in EEG signals by using the generalized extreme value distribution. The underlying idea is to estimate the maximum and minimum values of the signal in the central motor cortex using statistical modeling, for motor imagery events,  corresponding to mu-suppression. To the best of our knowledge, this statistical model has not been investigated yet for detecting the mu-suppression in EEG signals, despite the extensive study of this phenomenon \cite{Pineda2005,Hobson2017,Nishimura12018}. Several other methods have been proposed in the literature to estimate mu-suppression in motor imagery, see \cite{Hamedi2016,Xu2019} for a comprehensive state-of-the-art. Table \ref{tab:sart} summarizes the most common methods, such as independent components analysis, common spatial patterns, linear discriminant analysis, and the support vector machines.

\begin{table}[!htb]
\centering
\begin{tabular}{ | p{3.3cm} | p{5.2cm} | p{4cm} | p{0.6cm} |  p{0.6cm} |  p{0.6cm} |}
	\hline
	Method & Features & Electrodes & Classf.  & Perf. & Ref.\\
	\hline
	GEV from periodogram & GEV parameters & C5, C3, C1, Cz, C2, C4, C6 & LDA & 100\%  & \hlr{our}\\
	\hline
	AAR + BP  & Coefficients from AAR and BP & C3, Cz & LDA & 80\%& \cite{Guger2003}\\
	\hline
	RFE &  Welch method from RFE & C1, C2, C3, C4, FC3, FC4, CP3, CP4& SVM & 87\%& \cite{Schroder2005}\\
	\hline
	ICA vs.  CSP & log-transformed normalized variances& C3, C4, Cz, CP1, CP2, CPz, Fz, Pz & LDA & 84\% &\cite{Naeem2006}\\
	\hline	
	CSP & Eigenvalue spectrum & multi-channels& LDA & 84\% & \cite{Blankertz2008}\\
	\hline	
	EMD & Periodograms fluctuations from intrinsic mode functions and the Hilbert evelope &C3, C4& NR & NR &\cite{Wan2009}\\
	\hline
	SCSP & Variance of the spatially filtered signals & C3, C4, Cz & SVM & 94\%&\cite{Arvaneh2011}\\
	\hline
	ICA  & time-frequency average & multi-channels & LAT&90\%&\cite{Xu2019}\\
	\hline
\end{tabular}
\caption{State-of-the-art methods to the mu-suppression estimation in EEG signals, compared in terms classification techniques, features, and  reported performance. Abbreviations are as follows:  Adaptive autoregressive model (AAR), Band power  estimation (BP), Recursive feature elimination (RFE),  Independent components analysis (ICA),  Common spatial patterns (CSP), Empirical mode decomposition (EMD), Sparse common spatial pattern (SCSP), Linear discriminant analysis (LDA), Support Vector Machine (SVM), According to the latency (LAT), Not reported (NR). Figure \ref{fig:eeglocs} shows channel locations. Note that our method (highlighted in red) had excellent accuracy results.}
\label{tab:sart}
\end{table}

The remainder of the paper is organized as follows. In Section \ref{sec:gev}, the generalized extreme value distribution is presented. Section \ref{sec:mm} describes the database used and the proposed methodology. Experimentation using the EEG dataset of motor imagery from \cite{Cho2017} is presented in Section \ref{sec:res} and the results discussed. Finally, Section \ref{sec:con} draws conclusions and future works.

\section{Materials and Methods}
\label{sec:mm}

\subsection{Motor Imagery database}
\label{ssec:db}
\begin{figure}[!ht]
	\centering
	\subfigure[Central motor cortex]{\includegraphics[clip,width=0.35\columnwidth]{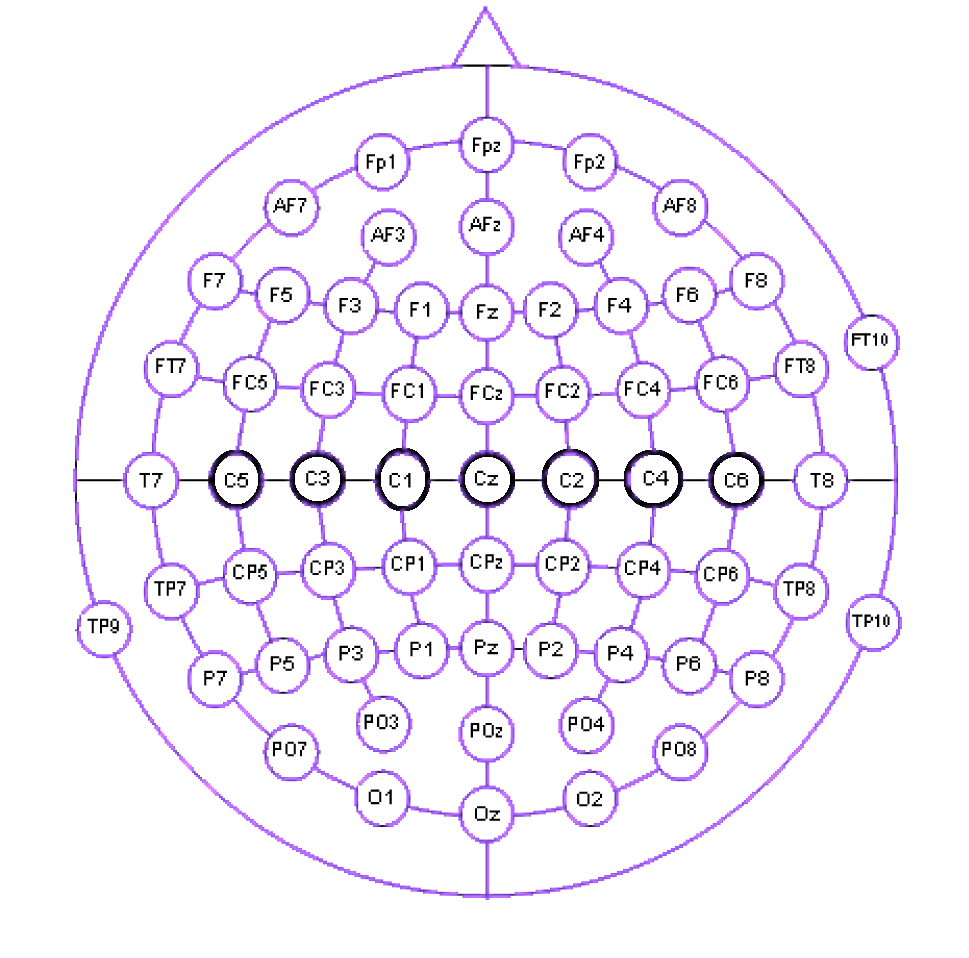}\label{fig:eeglocs}}
	\subfigure[Channel Cz]{\includegraphics[clip,width=0.6\columnwidth]{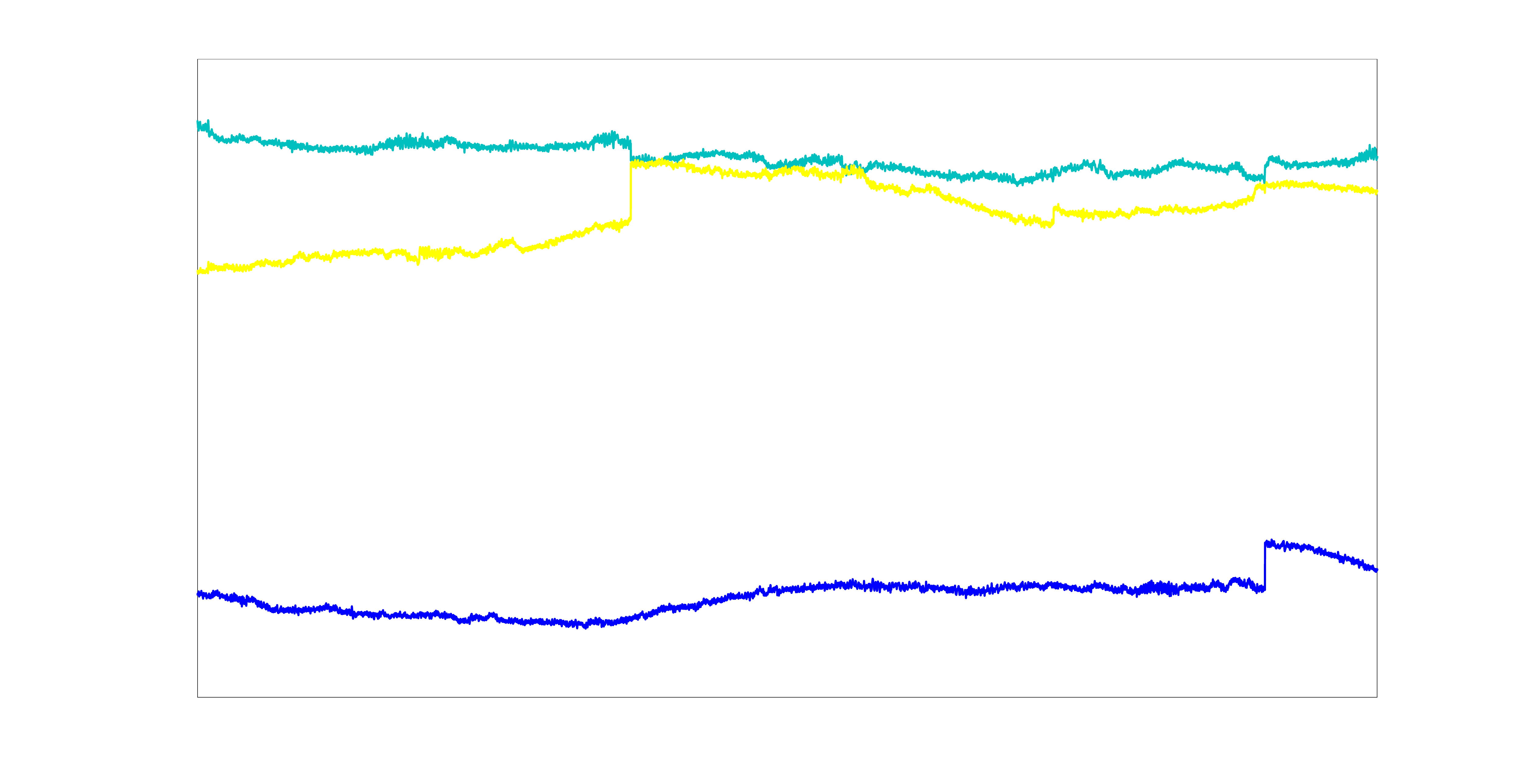}\label{fig:cz}}
	\caption{Figure (a) shows the central motor cortex channels (C5,C3,C1,Cz,C2,C4,C6) following the international 10-10 system. Figure (b) shows an example of the central channel for each type of event. Imagery event (grey) having the highest amplitude, movement event (yellow) with a medium amplitude, and resting event (blue) corresponds to the lower amplitude. Note that, imagery signals and movement signals may have very close amplitudes.}
\end{figure}

A motor imagery EEG database of the left and right hands taken from \cite{Cho2017} was used. EEG data of 52 healthy subjects were recorded using a 64 channel montage based on the international 10-10 system, with a 512 Hz sampling rate. In this database, 3 events were taken into account. First, non-task, with the motor cortex, data was recorded in order to understand each subject's performance related to their eye blinking, eyeball up/down movement, eyeball left/right movement, head movement, jaw clenching, and resting state. Second, real right and left-hand movement was measured, where subjects had to follow certain established patterns. Finally, motor imagery data was computed when asking subjects to imagine hand movement. The signals were preprocessed with a high pass and low pass Butterworth filter of order 4$^{th}$,  having an 8-30 Hz bandpass. Data was validated by using the percentage of bad trials, event-related desynchronization/synchronization (ERD/ERS) analysis, and classification analysis. The central motor cortex was selected for this study, see Fig. \ref{fig:eeglocs}.

\subsection{Statistical modeling}
\label{sec:gev}
We propose to model the periodogram power spectral density of the EEG data using the generalized extreme value distribution (GEV). The choice of this distribution is motivated by its flexibility. It is a family that combines the Gumbel (type I), Fr\'echet (type II) and Weibull (type III) maximum extreme value distributions through three parameters, namely location ($\mu \in \mathcal{R}$), scale ($\sigma > 0$), and shape ($\xi \in \mathcal{R}$ ). Its probability density function (PDF) is given by
\begin{align}
    \label{eq:pdf}
    f(x,\mu,\sigma,\xi) & = \frac{1}{\sigma}\;t(x)^{\xi + 1}\exp^{-t(x)}
\end{align}
where
\begin{align}
\label{eq:def}
    t(x) & = 
\begin{cases}
(1 + \xi(\frac{x-\mu}{\sigma}))^{-1/\xi} & \text{, if} \ \xi \neq 0
\\
\exp^{\frac{x-\mu}{\sigma}} & \text{, if} \ \xi = 0
\end{cases}
\end{align}
 $\xi$ determines the domain of the PDF as follows:
\begin{align}
    \label{eq:int}
    x \in \left[\frac{\mu-\sigma}{\xi},+\infty\right) &\text{, when } \xi>0 \text{, type II}\\
    \nonumber
    x \in \left(-\infty,+\infty\right) &\text{, when } \xi=0 \text{, type I}\\
    \nonumber
    x \in \left(-\infty,\frac{\mu-\sigma}{\xi}\right] &\text{, when } \xi<0 \text{, type III}
\end{align}
In order to fit the GEV distribution to the data, we calculated the parameters $\mu$ and $\sigma$  using the maximum likelihood estimators \cite{GoF1986}
\begin{align}
	\label{likelihood}
	\hat{\mu} &= -\hat{\sigma}\log \left( \frac{1}{n} \sum_{i=1}^n \exp^{\frac{X_i}{\hat{\sigma}}}\right) \\
	\hat{\sigma} &= \bar{X}-\frac{\sum_{i=1}^n X_i\exp^{-\frac{X_i}{\hat{\sigma}}}}{\sum_{i=1}^n \exp^{-\frac{X_i}{\hat{\sigma}}}}
\end{align}

The cumulative density function (CDF) is given by
\begin{align}
    \label{eq:cdf}
    f(x,\mu,\sigma,\xi) & = \exp^{-t(x)}
\end{align}
for $x \in$ type II  and type III.

We refer the reader to \cite{KotzNadarajah2001} for a comprehensive treatment of the mathematical properties of the extreme value distribution.
\subsection{Proposed methodology}
\label{ssec:meth}
Let $\bs X \in \mathbb{R}^{N\times M}$ denote the matrix of
 $M$ EEG signals $\boldsymbol{x}_m \in \mathbb{R}^{N\times 1}$ measured simultaneously on different channels related to the central motor cortex and at $N$ discrete-time instants. 
 The proposed methodology is composed of three stages. In the first-stage, the periodogram power spectral density (PSD) is estimated from the EEG signal on the mu-brain rhythm, in steps of half the sampling frequency without overlapping using a symmetric Hamming window
\begin{align}
\label{psd1}
	P(f)=\frac{\Delta t}{N}\left | \sum_{n=0}^{N-1} x_{m,n} h(n) \exp^{-2j\pi f \Delta t \;n} \right |^2,\-\frac{1}{2}\Delta t < f  \leq \frac{1}{2}\Delta t 
\end{align}
where $\Delta t$ is the sampling interval, $1 \leq m \leq 7$ is related to the seven channels in the central motor cortex (C5, C3, C1, Cz, C2, C4, C6) and the Hamming window defined as $h(n)=0.54 -0.46 \;cos \left ( 2\pi \frac{n}{N}\right)$, $0\leq n\leq N$. 

In the second-stage, the generalized extreme value distribution is fitted to the PSD of each event (imagery, movement and resting), yielding the corresponding  parameters ($\mu$, $\sigma$ and $\xi$):
\begin{align}
	\label{eq:ima}
	I_i&=[\mu_i, \sigma_i, \xi_i]\\
	\label{eq:mov}
	M_i&=[\mu_i, \sigma_i, \xi_i]\\
	\label{eq:res}
	R_i&=[\mu_i, \sigma_i, \xi_i]
\end{align}
Finally in the third stage, the vectors \eqref{eq:ima},\eqref{eq:mov} and \eqref{eq:res} are classified using a linear classifier derived from a linear discriminant analysis (LDA)\cite{QuinteroRincon2018a,QuinteroRincon2018b}. Three  classes $\{0, 1, -1\}$ are possible, respectively for imagery, movement and resting.

\section{Results and discussion}
\label{sec:res}

\begin{figure}[!ht]
	\centering
	\subfigure{\includegraphics[clip,width=1\columnwidth]{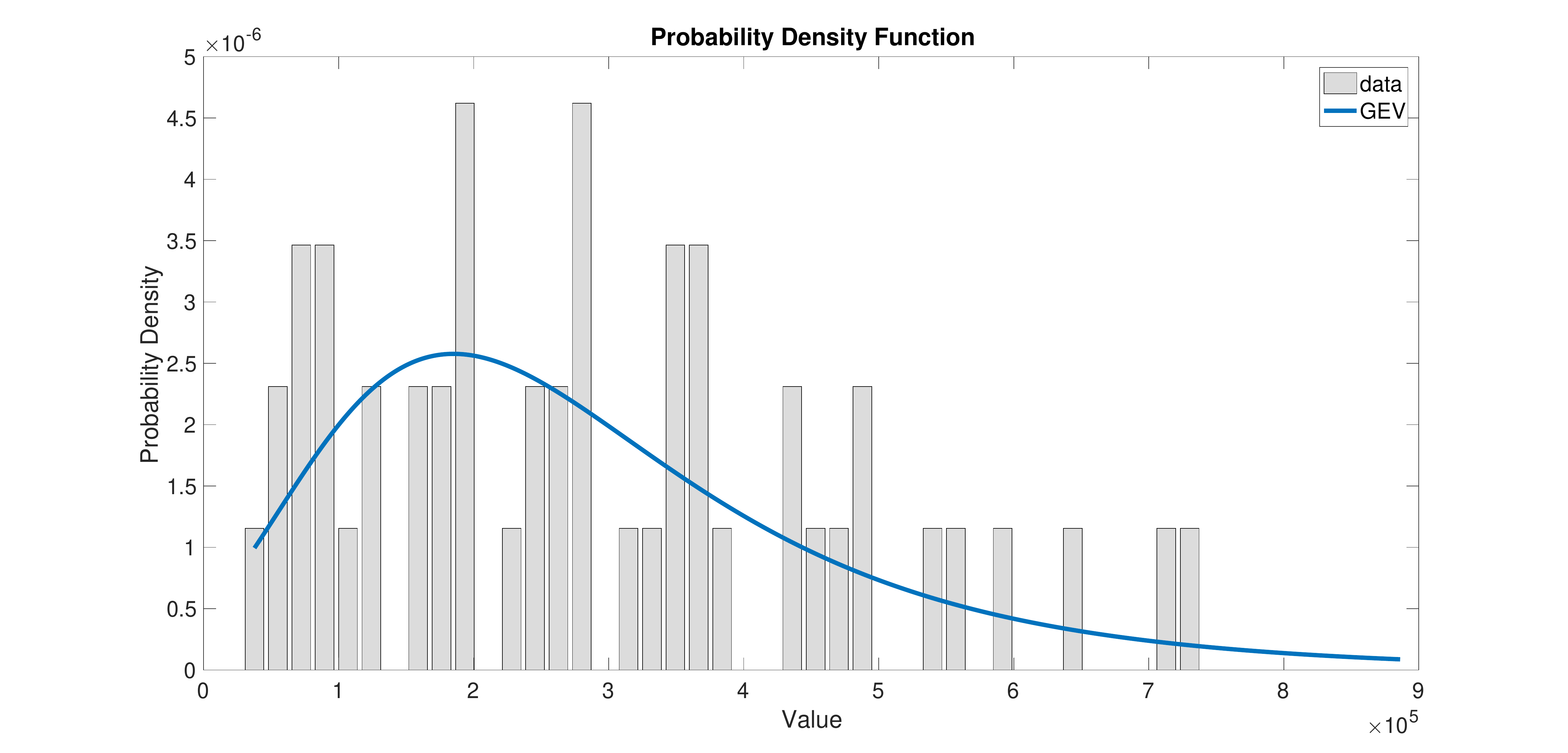}}
	\subfigure{\includegraphics[clip,width=1\columnwidth]{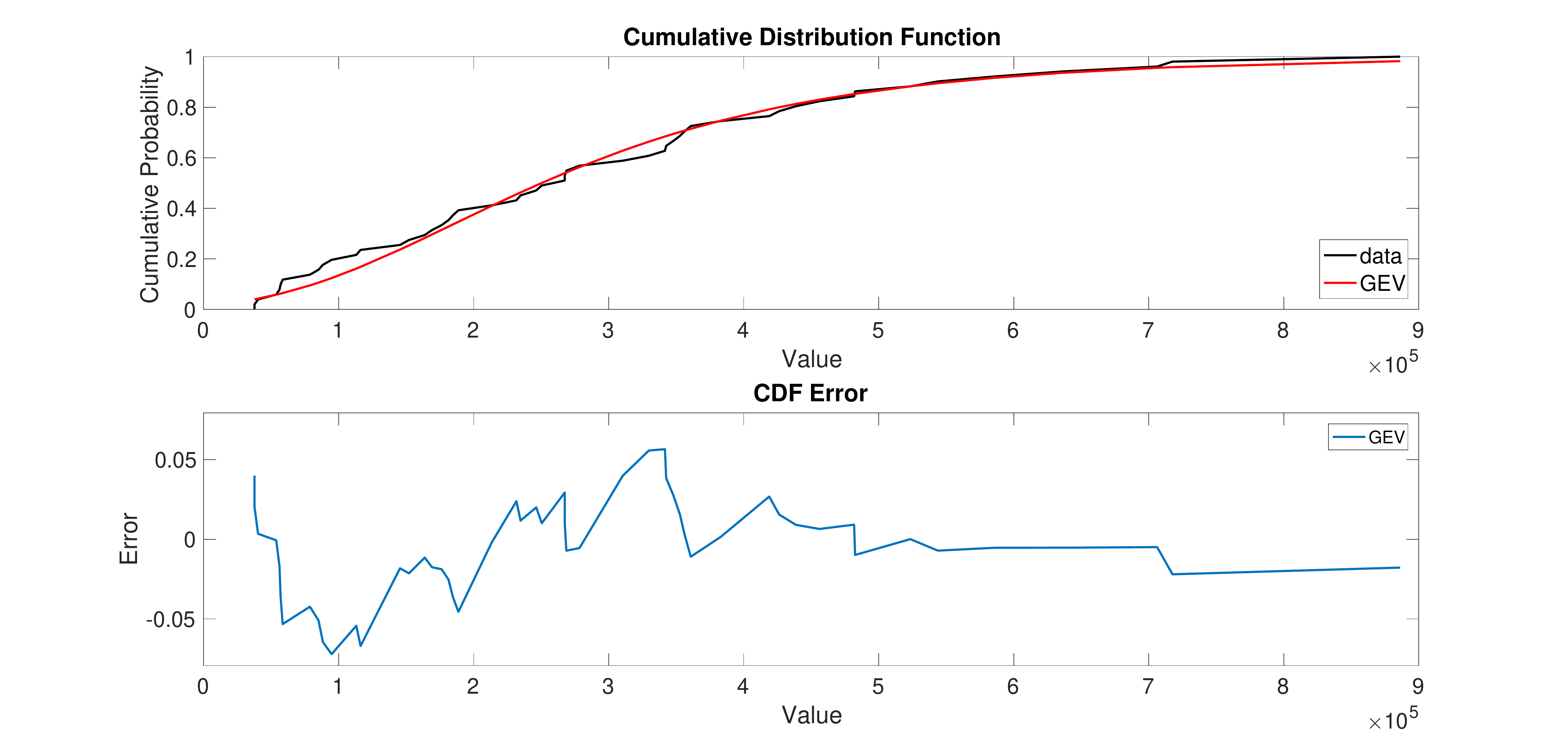}}
	\caption{Mu-band in imagery event}
	\label{fig:imag}
\end{figure}

\begin{figure}[!ht]
	\centering
	\subfigure{\includegraphics[clip,width=1\columnwidth]{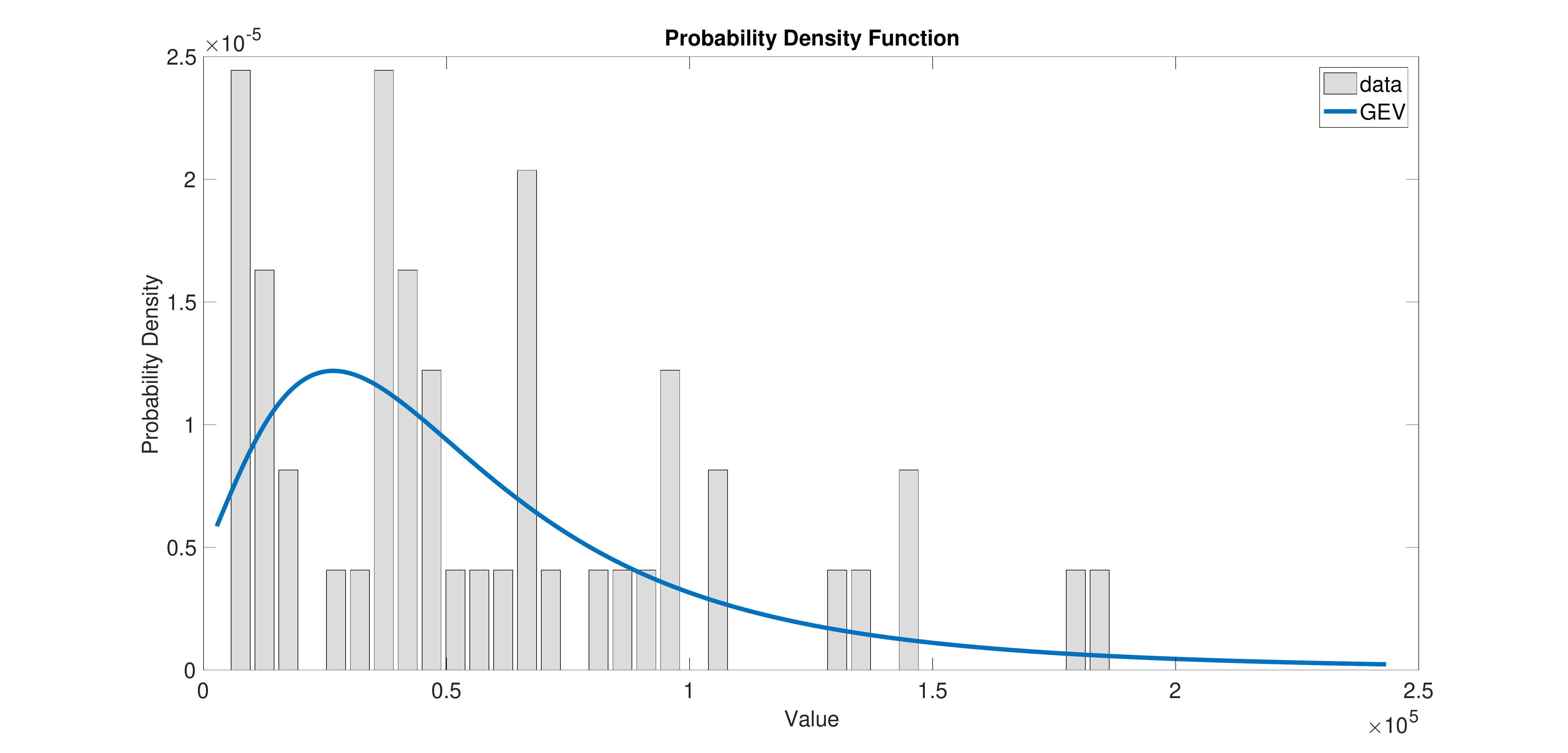}}
	\subfigure{\includegraphics[clip,width=1\columnwidth]{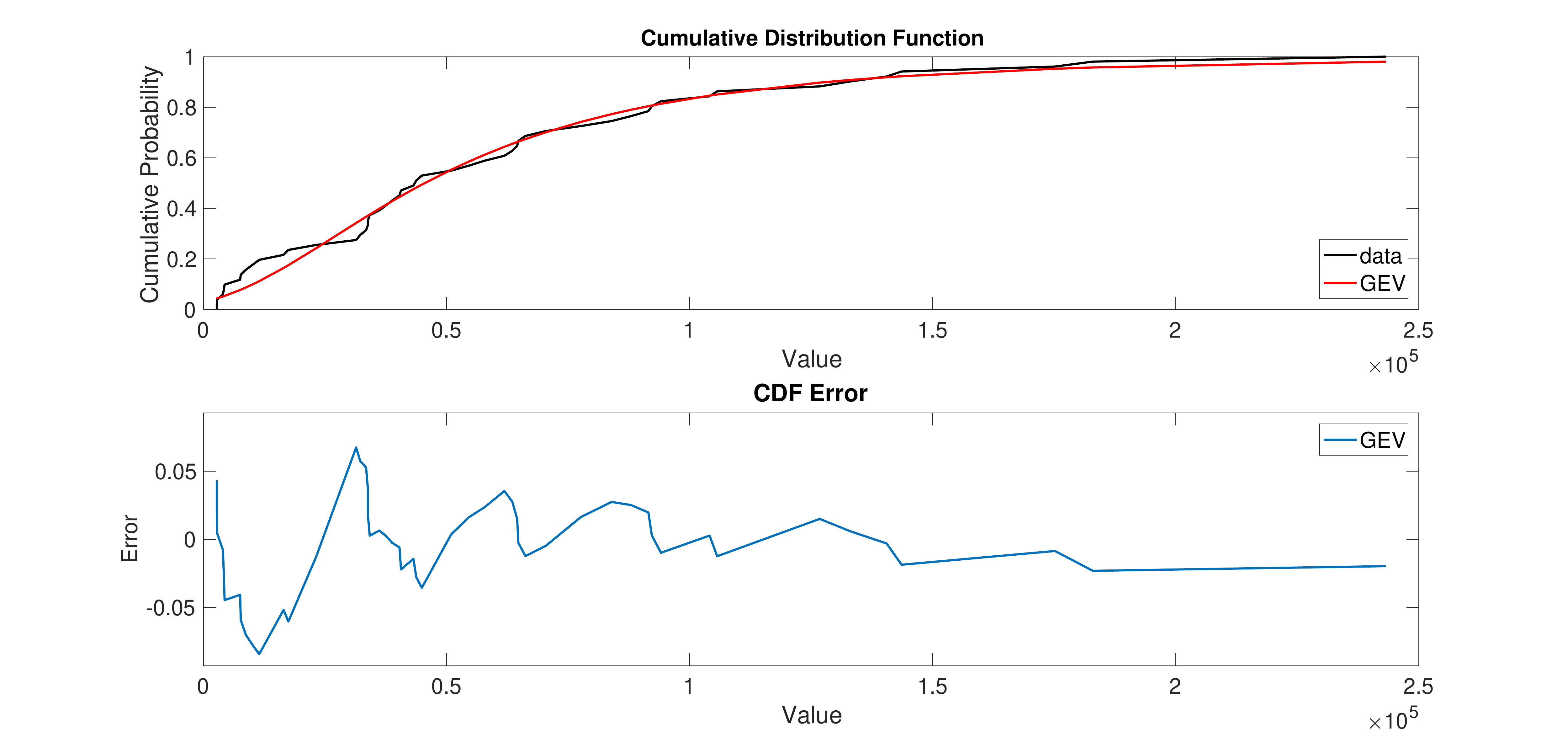}}
	\caption{Mu-band in movement event}
	\label{fig:move}
\end{figure}

\begin{figure}[!ht]
	\centering
	\subfigure[Resting]{\includegraphics[clip,width=1\columnwidth]{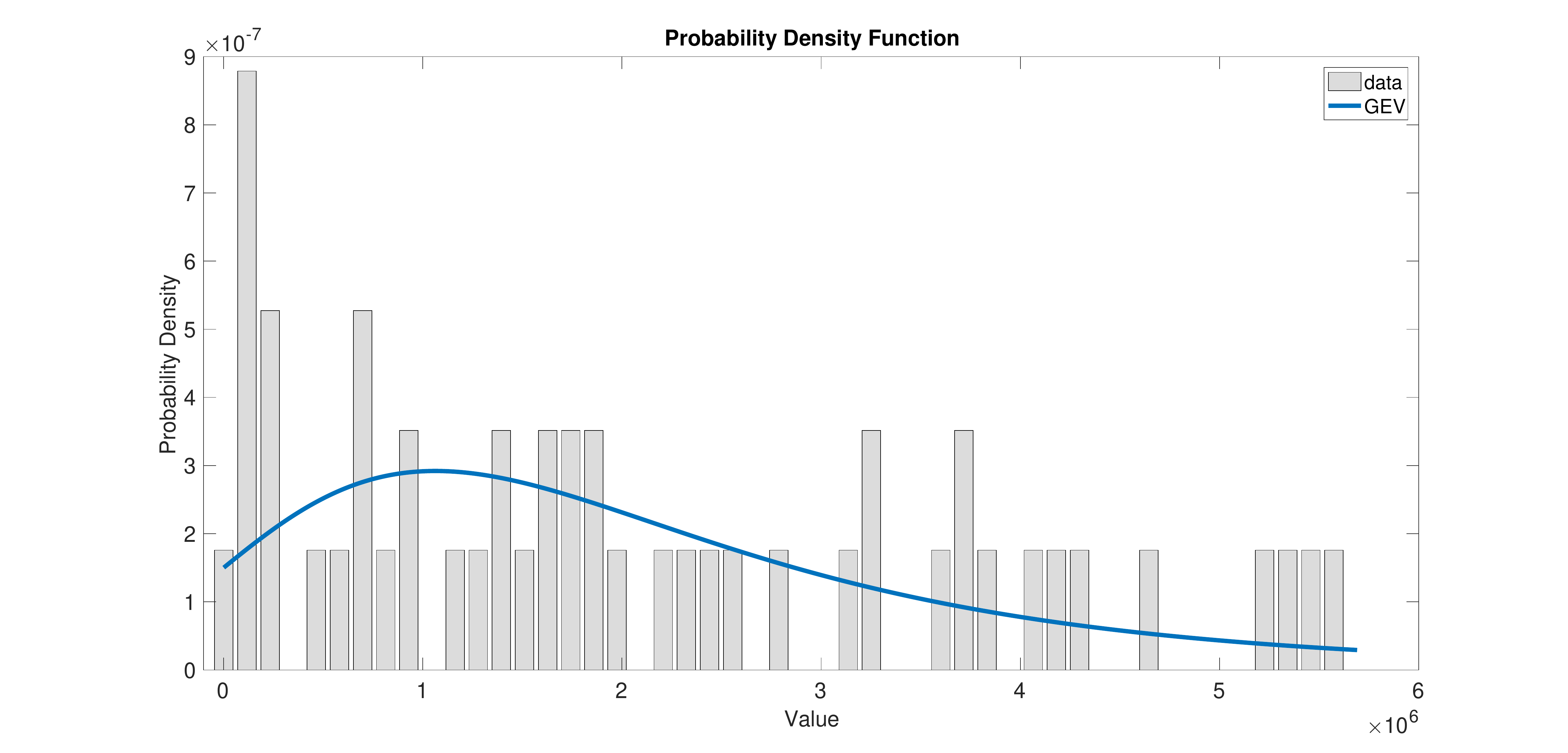}}
	\subfigure[Resting]{\includegraphics[clip,width=1\columnwidth]{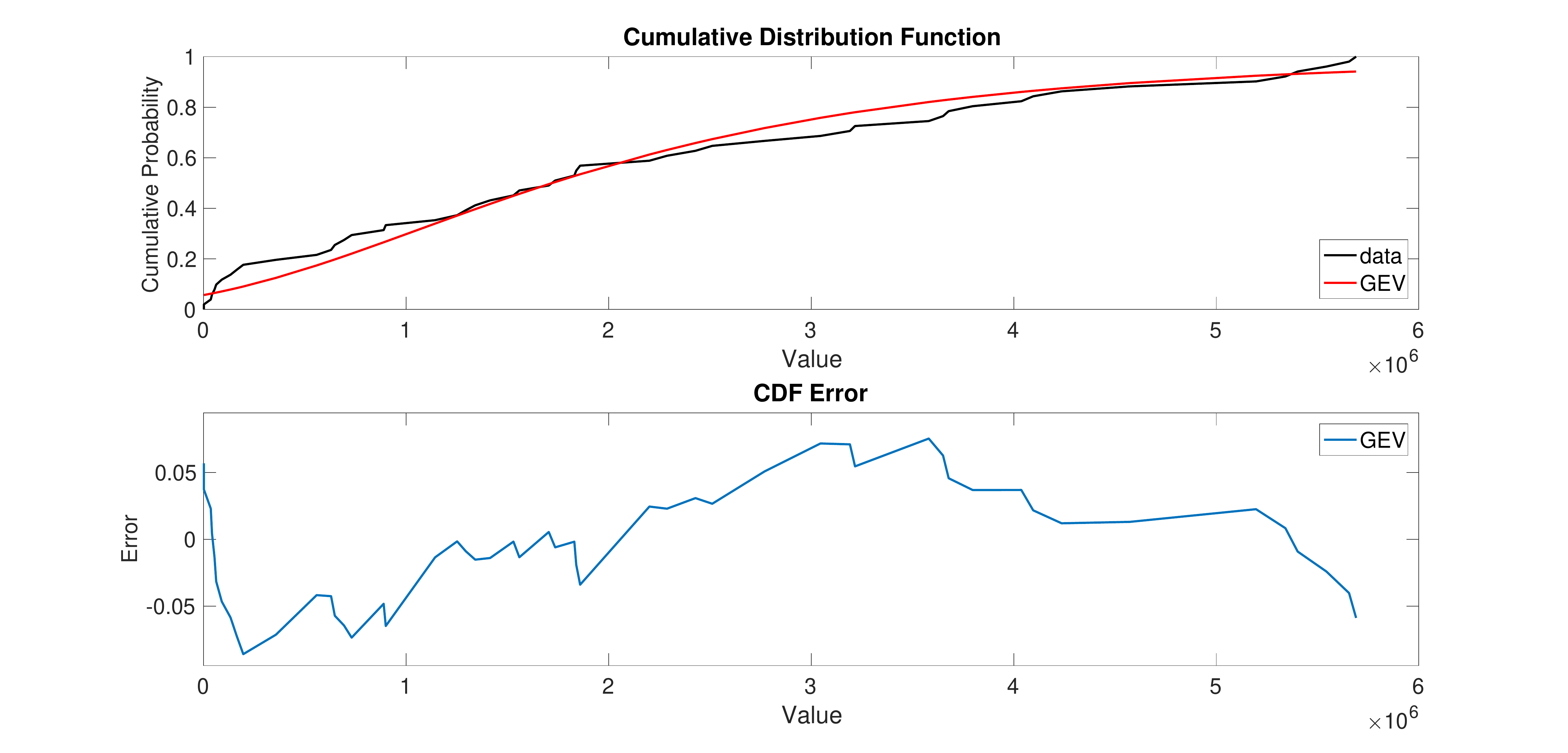}}
	\caption{Mu-band in resting event}
	\label{fig:rest}
\end{figure}

Figures \ref{fig:imag}, \ref{fig:move}, and \ref{fig:rest} show the Goodness-of-fit measures for the probability density function (PDF) and the cumulative distribution density (CDF), and the CDF error for the central motor cortex data. These measures are clearly significant. This is corroborated by the CDFs errors ($-0.05 \leq \text{CDF error} \leq 0.05$), and the $p$-values $\leq 0.05$ using the Kolmogorov-Smirnov (KS) score, where the imagery events exhibit a $p= 0.0021$, the movement events a $p=0.0094$, and the resting events a $p=1.9847e-04$.

\begin{figure}[!ht]
	\centering
	\includegraphics[clip,width=1\columnwidth]{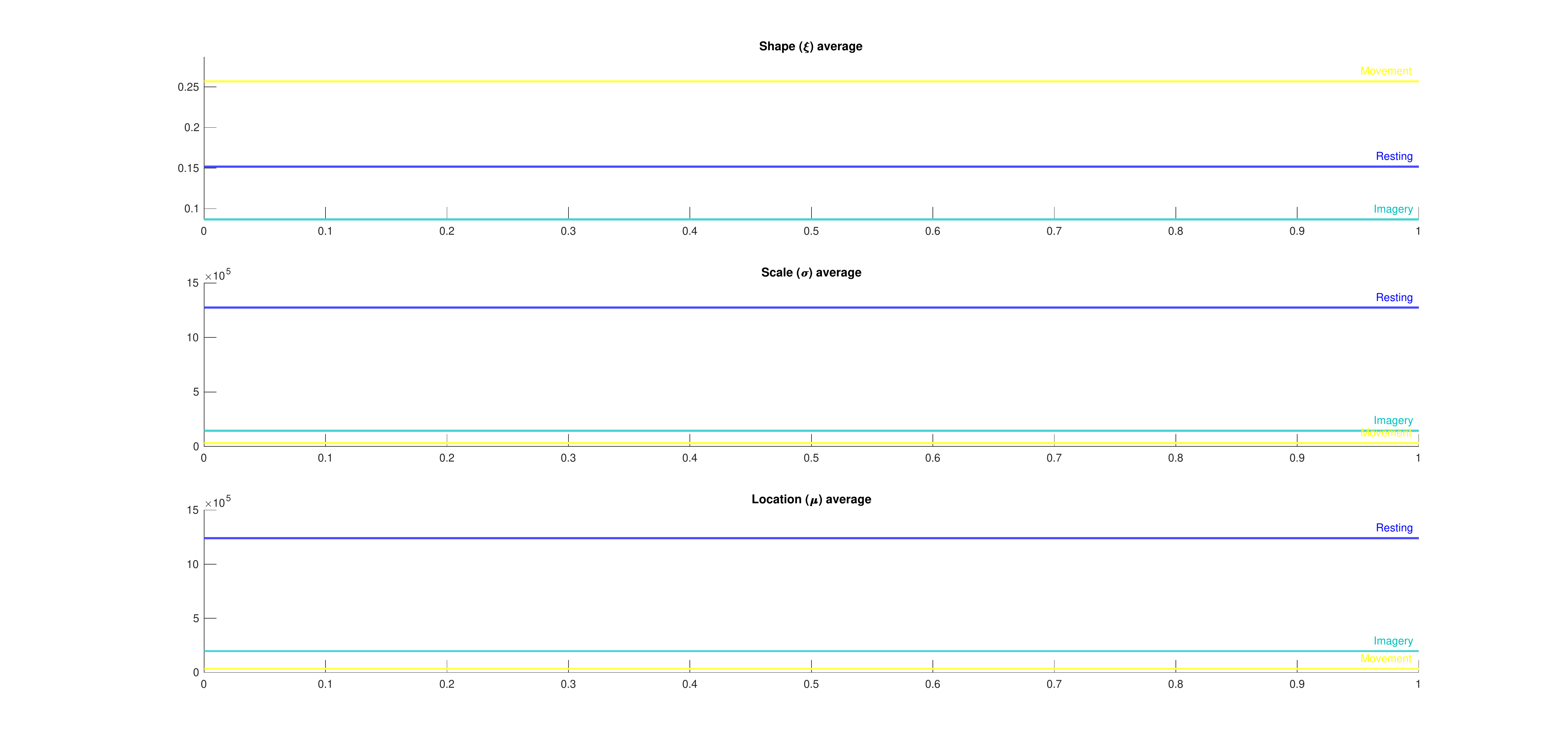}
	\caption{Plot of the mean values from Tables \ref{tab:gevcentral1} to \ref{tab:gevcentral3}. Note that, these parameters allow the identification of imagery events (grey), the movement events (yellow), and the resting events (blue) using a threshold approach.}
	\label{fig:means}
\end{figure}

Tables \ref{tab:gevcentral1} to \ref{tab:gevcentral3} show the mean values, with their $95\%$ confidence intervals, for the three GEV parameters corresponding to the different events. It appears clearly that the proposed statistical model allows for a threshold approach to detect the mu-suppression by distinguishing  imagery, from movement and resting events.  For illustration, in EEG data shown in Figure \ref{fig:cz}, the imagery event has the highest amplitude, the movement event has the medium amplitude, and the resting event has a lower amplitude. While in Figure \ref{fig:means}, showing averages from Table  \ref{tab:gevcentral1}, the scale ($\sigma$) and the location ($\mu$) parameters have the highest amplitude for resting state,  the medium amplitude for imagery event, and the lower amplitude for movement event. Whereas, the shape ($\xi$) parameter shows the movement event with the highest amplitude, the resting event with the medium amplitude and the imagery state with the lower amplitude. The combination  of these different thresholds makes it  possible to use the three parameters, estimated by fitting the GEV distribution to the periodogram, for detecting the mu-suppression in motor imagery EEG signals.

\begin{table}[!ht]
\centering	
\begin{tabular}{|c|c|}%
	\hline 
	& shape ($\xi$)  \\ 
	\hline 
	Ima     & 0.09 [-0.21,0.38] \\
    	\hline
   	 Mov   & 0.26 [-0.06,0.58]  \\
    	\hline
    	Res    & 0.15 [-0.26,0.56]   \\
\hline
 \end{tabular} 
\caption{Mean values of the shape parameter of the GEV adjusted to the periodogram of mu-band, with the 95\% confidence intervals.}
\label{tab:gevcentral1}
\end{table}
\begin{table}[!ht]
\centering	
\begin{tabular}{|l|r r r|}%
	\hline 
	& \multicolumn{3}{c|}{scale ($\sigma$)} \\ 
	\hline 
	Ima     &143275.14 &[110712.36,&185415.39]  \\
    	\hline
   	 Mov   &31104.66 & [23545.71,&41090.27]  \\
    	\hline
    	Res    & 1273712.63 & [945447.56,&1715953.13]  \\
\hline
 \end{tabular}
\caption{Mean values of the scale parameter of the GEV adjusted to the periodogram of mu-band, with the 95\% confidence intervals}.
\label{tab:gevcentral2}
\end{table}
\begin{table}[!ht]
\centering	
\begin{tabular}{|l|r r r|}%
	\hline 
	& \multicolumn{3}{c|}{location ($\mu$)}\\ 
	\hline 
	Ima    & 196975.08 &[149574.66,&244375.50] \\
    	\hline
   	 Mov   & 33551.02 &[23148.30,&43953.76]  \\
    	\hline
    	Res    & 1241060.75 &[778886.63,&1703234.88] \\
\hline
 \end{tabular} 
\caption{Mean values of the location parameter of the GEV adjusted to the periodogram of mu-band, with the 95\% confidence intervals}.
\label{tab:gevcentral3}
\end{table}


To assess the quality of the proposed method for mu-band suppression, we adopted a supervised classification approach with 30 thousand randomized signals for training and testing from the motor imagery database with the 3 GEV features for each vector as follows:

\vspace*{2em}

\begin{itemize}
	\item $\theta_1=[\mu_1, \sigma_1, \xi_1]$  10 thousand randomized imagery events.
	\item $\theta_0=[\mu_0, \sigma_0, \xi_0]$ 10 thousand randomized movement events.
	\item $\theta_{-1}=[\mu_{-1}, \sigma_{-1}, \xi_{-1}]$ 10 thousand randomized resting events.
\end{itemize}

A linear classifier derived from the linear discriminant analysis (LDA) was designed to distinguish between the different events $[\theta_1, \theta_0, \theta_{-1}]$. Note that, the LDA computational complexity is lower with respect to support vector machines (SVM), which is one of the most used classifiers. The LDA computational complexity is $O(mn+mt+nt)$, where $m$ is the number of samples, $n$ is the number of features, and $t=min(m,n)$. While SVM complexity is $O(n^2)$ or $O(n^3)$ depending on the parametrization, with $n$ being the size of the dataset, which is generally large when EEG signals are analyzing \cite{QuinteroRincon2018b}.

\begin{figure}[!h]
\centering
\includegraphics[clip,width=1\columnwidth]{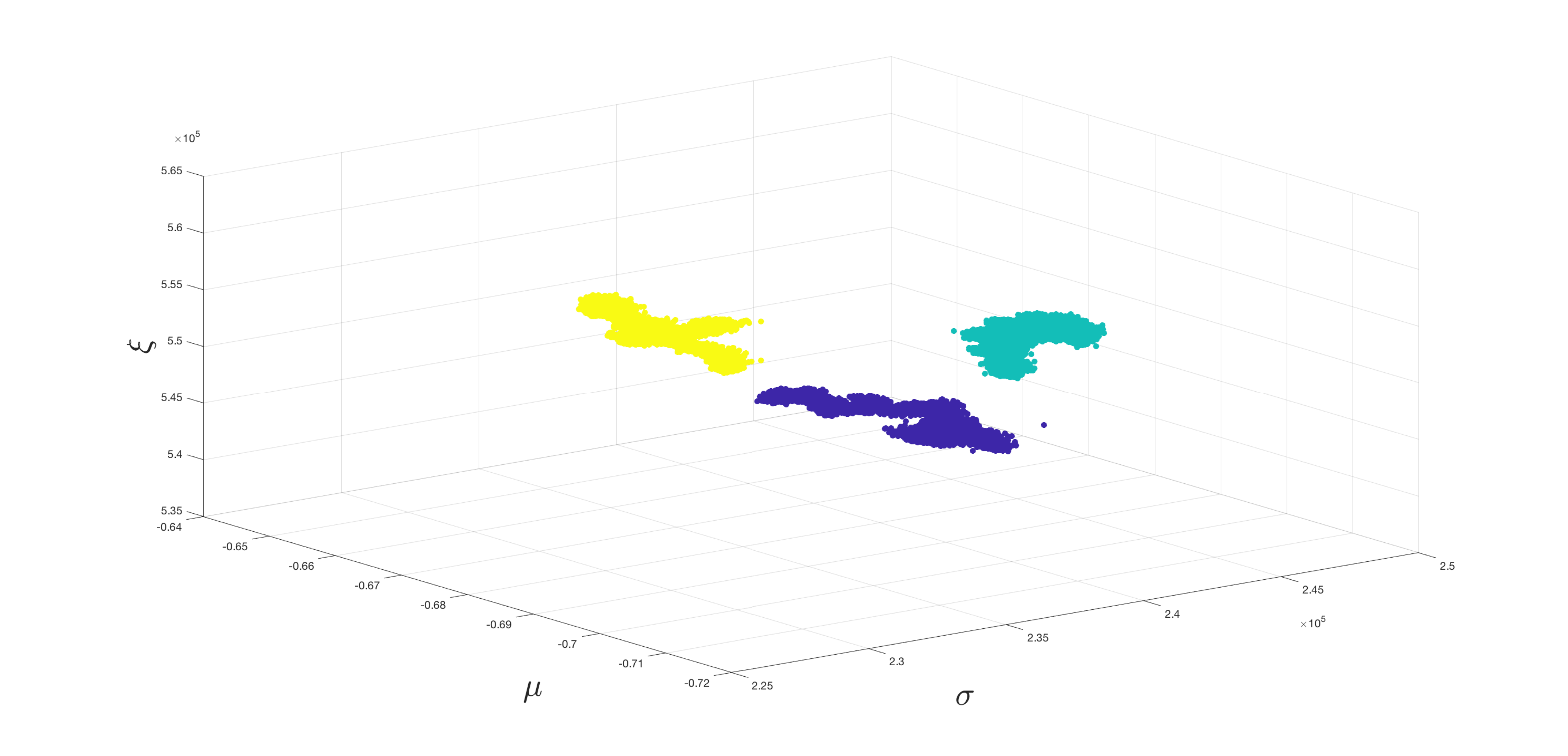}
\caption{Scatter plot example with randomized data. It is clearly possible to use a linear classifier as imagery events (grey), the movement events (yellow), and the resting events (blue) are well separated in the GEV parameter space.}
\label{fig:gev}
\end{figure}

Due to the big data volume generated, a 20-fold cross-validation technique was used for training and testing, with 10 times repetition to ensure no bias in the partitioning and to avoid imbalanced data \cite{QuinteroRincon2019d}.  Figure \ref{fig:gev}, shows the scatter distribution of the different randomized events, namely imagery, movement, and resting. It is interesting to note the separability of the classes. This justifies evidently the use of linear classification to distinguish events. 
The classifier achieves a 100\%  sensitivity (True positive rate) and specificity (True false rate) for mu-suppression detection.


\section{Conclusions}
\label{sec:con}
This work presented a study of detecting mu-suppression in EEG signals, based on the generalized extreme value (GEV) distribution. The three parameters of this distribution were estimated from its periodogram power spectral density for three events: imagery, movement, and resting. A linear classifier derived from the linear discriminant analysis (LDA) was designed to distinguish between event types using these parameters. The performance of the proposed method was evaluated on a real dataset from 52 subjects achieving 100\% accuracy. In addition to its performance, an advantage of this method is its low computational cost compared to existing methods. These good results have the potential to shed new light on mu-suppression detection in motor imagery EEG signals in the central motor cortex.

The noise and artifacts were not taken into consideration in this work, which constitutes its main limitation. Future work will focus on the study of the noise and artifacts in order to make the method applicable in real-time. A large scale testing campaign will also be undertaken with other databases and other possible channel locations with the idea of using the least amount of channels possible.

\section{Acknowledgments}
The authors would like to thank Joaquin Ems, Lourdes Hirschson, and Catalina Carenzo for useful comments on an earlier version of the manuscript. We are grateful to the professor emeritus Jaime A. Pineda from the University of California, San Diego, La Jolla, CA, USA, for all the useful discussions during the process of this study.

\bibliographystyle{unsrt}

\end{document}